\documentclass[onecolumn,showpacs,preprintnumbers,amsmath,amssymb,eqsecnum]{revtex4}
\usepackage{epsfig}
\usepackage{graphicx}
\usepackage{rotating}

\begin{document}
\title{Cosmological measure with volume averaging and the vacuum energy problem}
\author{Artyom V. Astashenok${}^{1}$ and Antonino Del Popolo${}^{2,3,4,5}$\\
\small ${}^{1}$I. Kant Baltic Federal University\\
\small Department of Theoretical Physics\\
\small 14, Nevsky st., 236041 Kaliningrad, Russia\\
\small ${}^{2}$ Dipartimento di Fisica e Astronomia, Universit$\acute{a}$ a di Catania\\
\small Viale Andrea Doria 6, 95125 Catania, Italy\\
\small ${}^{3}$ Instituto de Astronomia, Geof\'{\i}sica e Ci\^encias Atmosf\'ericas \\
USP, 05508-900 S\~ao Paulo, SP, Brasil\\
\small ${}^{4}$ Astronomical Observatory, Kyiv National University\\
3 Observatorna St., 04053 Kyiv, Ukraine\\
\small ${}^{5}$ Main Astronomical Observatory, Academy of Sciences of Ukraine\\
\small 27 Akademika Zabolotnoho St., 03680 Kyiv, Ukraine}

\begin{abstract}

In this paper, we give a possible solution to the cosmological constant problem. It is shown that the traditional approach, based on volume weighting of probabilities, leads to an incoherent conclusion: the probability that a randomly chosen
observer measures $\Lambda=0$ is exactly equal to 1. Using an alternative, volume averaging measure, {instead of volume weighting} can explain why the cosmological constant is non-zero.

\end{abstract}



\maketitle

\section{Introduction}

The problem of vacuum energy is probably one of the most interesting puzzles of modern physics. The observable energy density of the vacuum is at least 60 orders of magnitude smaller {than the value expected from the Standard Model}. {There is no known natural way to derive the tiny cosmological constant used in cosmology from particle physics.}

These problem can be explained through the anthropic principle. In a well-known article \cite{Weinberg}, Weinberg estimated the upper limit for the effective cosmological constant as
$$
\Lambda_{max}<5000\Lambda_{0},
$$
where $\Lambda_{0}$ is an observable value. Higher values suppress hierarchical structure formation in the universe, and, therefore, lead to cosmologies completely devoid of life as we know it.

A further development in the application of anthropic argument was the inclusion of selection rules such as a self-sampling assumption or "mediocrity principle", namely  {the notion that there is nothing very unusual about our civilization.} Acceptance of mediocrity principle, grounded on statistical approach, explicitly implies the existence of a multiverse that serves a role of a statistical ensemble. {In multiverse, we can estimate the probabilities of observing any given event $j$.} Such a probability is factorized as
\begin{equation}
P_j\sim {\bar{P}}_jf_j,
\label{factor}
\end{equation}
where ${\bar{P}}_j$ is the concentration of j-type bubbles and $f_j$ is the anthropic factor proportional to a total amount of observers residing inside the j-type bubble.

At first glance it may seem that the approach based on (\ref{factor}), could solve the problem of cosmological constant. Indeed, the anthropic factor decreases with increasing $\Lambda$. However, there is another difficulty. As we shall see, when $\Lambda \rightarrow 0$ the number of observers within a bubble increases and tends to infinity. This phenomenon may be called the "infrared divergence" \cite{Infra-1}, \cite{Infra}. Therefore the value $\Lambda=0$ is preferred from the anthropic point of view.

The main plane of article is the following. The next section is devoted to calculations of probabilities in multiverse according to (\ref{factor}). The existence of the "infrared divergence" is demonstrated. In the third section  we use another method to calculate the probabilities that the divergence doesn't appear. In conclusion possible objections are considered.

\section{Infrared divergence at $\Lambda\rightarrow 0$}

The probability $P_{\Lambda}$ to find oneself in a universe with a given value of vacuum energy can be determined via
\begin{equation}
P_{\Lambda}\sim N(\Lambda)\bar{P}(\Lambda),
\label{1}
\end{equation}
where $\bar{P}(\Lambda)$ is an a priori probability distribution (the relative abundance of different values of $\Lambda$ associated with the different types of bubbles in the multiverse), and $N(\Lambda)$ is an anthropic factor proportional to the total number of observers in a given region of a multiverse. The aforementioned number is evidently related to the star formation rate \cite{Vilenkin-0}, which can be estimated from the astrophysical data:
\begin{equation}
N(\Lambda)\sim \int^{t_{c}}_{0} \dot{n}(t, \Lambda)V_{c}(t)dt,
\label{2}
\end{equation}
where $\dot{n}(t, \Lambda)$ is the star formation rate in a comoving volume $V_{c}$ and $t_{c}$ defines a time of collapse. Obviously, for those universes whose expansion has a de Sitter-type asymptote $t_{c}=\infty$.

If the universe has zero curvature and the radiation is small enough, the Friedman equations become integrable and the scale factor may be written as
\begin{equation}
a=T^{2/3}\sinh^{2/3}{\tau}
\label{eq:8}
\end{equation}
where $T = 2/\sqrt{3 \Lambda}$ and $\tau$ is a dimensionless time $\tau=t/T$.

In order to calculate the comoving volume, $V_{c}$, we'll use the causal patch cut off. The causal patch is the region within the cosmological horizon \cite{Susskind-1}, \cite{Susskind-2}. Such a choice will result in
\begin{equation}
V_{c}(t)=\frac{4\pi}{3} \left(\int^{\infty}_{t}\frac{dt}{a(t)}\right)^{3}=\frac{4\pi T}{3} \left(\int^{\infty}_{\tau}\frac{d\tau}{\sinh^{2/3}(\tau)}\right)^{3}
\label{Vc}
\end{equation}
For $t\ll t_{\Lambda}=\sqrt{3/\Lambda}$ the scale factor changes according to a power law
$$
a(t)\approx t^{2/3}.
$$
When $\Lambda\rightarrow 0$, $V_{c}(t)$ diverges. The anthropic factor is inversely related to the cosmological constant. The resulting $P(\Lambda)$ will be determined by the previously chosen function $\bar{P}(\Lambda)$. The natural choice is a flat prior distribution i.e. $dP(\Lambda)/d\Lambda=0$, because the interval of anthropically acceptable values of $\Lambda$ is small in comparison with Planck scale.  {As we shall see, for such a distribution the probability to find oneself in a universe with smaller values of cosmological constant is larger.} Therefore the case $\Lambda=0$ is preferred.

\textbf{Remark}. One note that for $\Lambda=0$ Eq. (\ref{Vc}) is not applicable. In this case the infinity does not follow by formally taking the limit as $\Lambda\rightarrow0$. But for Friedmann universe without vacuum energy there is no cosmological horizon and therefore the comoving volume is infinite.

Let`s consider this fact in detail. The behavior of $\dot{n}(t)$ for our universe depends on the particular model of star formation \cite{Star-1},\cite{Star-2},\cite{Star-3}, \cite{Star-4}. However, the difference in $\dot{n}(t)$ using different star formation models
is not large, since all these models predict that $\dot{n}(t)$ reaches a maximum after a couple of billions of years and then is subject to a relatively fast decrease. The height and the width of the maximum depend on the cosmological constant.

Bousso and Leichenauer (hereafter BL), developed a semi-analytic model \cite{Star-5} of the star formation rate as a function of time, studying in particular how spatial curvature, amplitude of primordial density perturbations, and cosmological constant influence the SFR. Differently from previous papers (e.g., Hernquist \& Springel 2003, hereafter HS) their model is principally interested in how large changes in the studied parameters affect the SFR. HS model, for example is no longer valid when one studies SFR under large variations of the quoted parameters.

In order to understand the shape of the SFR, we have to recall two things:

a) structure formation originates from tiny perturbations already present in the early Universe. They expanded with Universe and collapsed before recombination giving rise to dark matter haloes that formed the gravitational wells in which baryons felt after recombination;

b) there was no star formation until structure formed with $T_{} > 10^4$ K. After structure formed, the SFR started to rise and reached a maximum.

Apparently there are significant differences between the results of BL model and HS and his calculations. One should point out that BL star formation model predicts that the rate at the present day is much smaller than in the other models. Also the other models seem to fit the data better at these late times. This is natural, since these models are adjusted to fit the data and BL model is not. This is also why the other two models closely agree with each other close to the present epoch.

A similar question arise concerning the  height and position of peaks in these  models. The top amplitude of SFR differs significantly: the biggest in BL model, the smallest for the fossil model \cite{Star-2}. In reality, the data about the peak of the star formation rate is much less certain \cite{Star-6}. One can see that the other models do not agree with each other concerning the epoch when the peak formed and its width. According to BL (private communication) the data, for large redshift, near the peak, was not as reliable as the data for small redshift and so it's much less clear which model is closer to reality.

The detailed calculations in \cite{Star-5} show that for sufficiently wide range of $\Lambda$, SFR varies only slightly (all other parameters remain fixed to the observed values; see especially figs. 3 and 4 in \cite{Star-5}).

One question at this point is why for large $\Lambda$ values, the SFR weakly depends on $\Lambda$ (e.g., $\Lambda=10\Lambda_{0}$), while for small $\Lambda$ ($0<\Lambda<\Lambda_{0}$), we have the same SFR, and the same total stellar mass  production per unit of comoving volume as in our universe.

A brief explanation of this issue is that in universe, star formation peaked around 3 Gyr and dropped off since then for a number of reasons.  At 3 Gyr the cosmological constant was dynamically unimportant and we might as well set it to zero.

Vacuum energy became important only several Gyr later, when star formation was already lower.  Without a cosmological constant, it is true that there would be more mergers in the future but these will mostly be halos that are too massive to cool efficiently.  So while $\Lambda$ does suppress hierarchical structure formation in the future, it has little effect on star formation, and decreasing $\Lambda$ would not significantly increase star formation in the future.

Therefore, when one is interested in numerical estimations, one can assume (for example in the range $0<\Lambda<10\Lambda_{0}$) that:
$$
\dot{n}(t, \Lambda)\approx \dot{n}(t, \Lambda_{0}),\quad 0<\Lambda<10\Lambda_{0}
$$
The accuracy of these estimates, in any case, is limited by our knowledge of star formation mechanisms. In the following, we consider the analytical fit for star formation rate in our universe given in \cite{Star-1}. The dependence of SFR from redshift $z$ is
\begin{equation}
\dot{n}(z)=\dot{n}_{m}
\frac{b\exp\left[a(z-z_m)\right]}{b-a+a\exp\left[b(z-z_m)\right]},
\label{eqnoldfit}
\end{equation}
with $a= 3/5$, $b=14/15$, $z_m=5.4$, and $\dot{n}_{0}= 0.15\,{\rm
M}_\odot{\rm yr}^{-1}{\rm Mpc}^{-3}$ is the maximal star formation rate at $z=z_{m}$.  {We plot this expression in Fig. 1(a), where the star formation rate reaches a peak at a redshift $z_{m}=5.4$, declining roughly exponentially towards both low and high redshift.} Using the well-known relation between $t$ and redshift $z$
$$
H_{0}dt=\frac{-dz}{((1+\Omega_{m}z)(1+z)^{2}-(z+2)z\Omega_{\Lambda})^{1/2}},
$$
where $H_{0}$ is the Hubble constant in our time and $\Omega_{m, \Lambda}=\rho_{m,\Lambda}/(\rho_{m}+\rho_{\Lambda})$, we can derive
the time dependence of $\dot{n}(t)$ {depicted in Fig. 1(b)}. We used $\Omega_\Lambda = 0.72\pm 0.04$ from the
WMAP results of \cite{WMAP} and $H_0 =72\pm 8$ km/s/Mpc from the Hubble Space Telescope key project \cite{HST}. {From Fig. 1b, one can see that star formation rate reaches maximum at $t_{m} \approx 1.5$ Gyr and then steadily declines.}

\begin{figure}
\begin{center}
\includegraphics{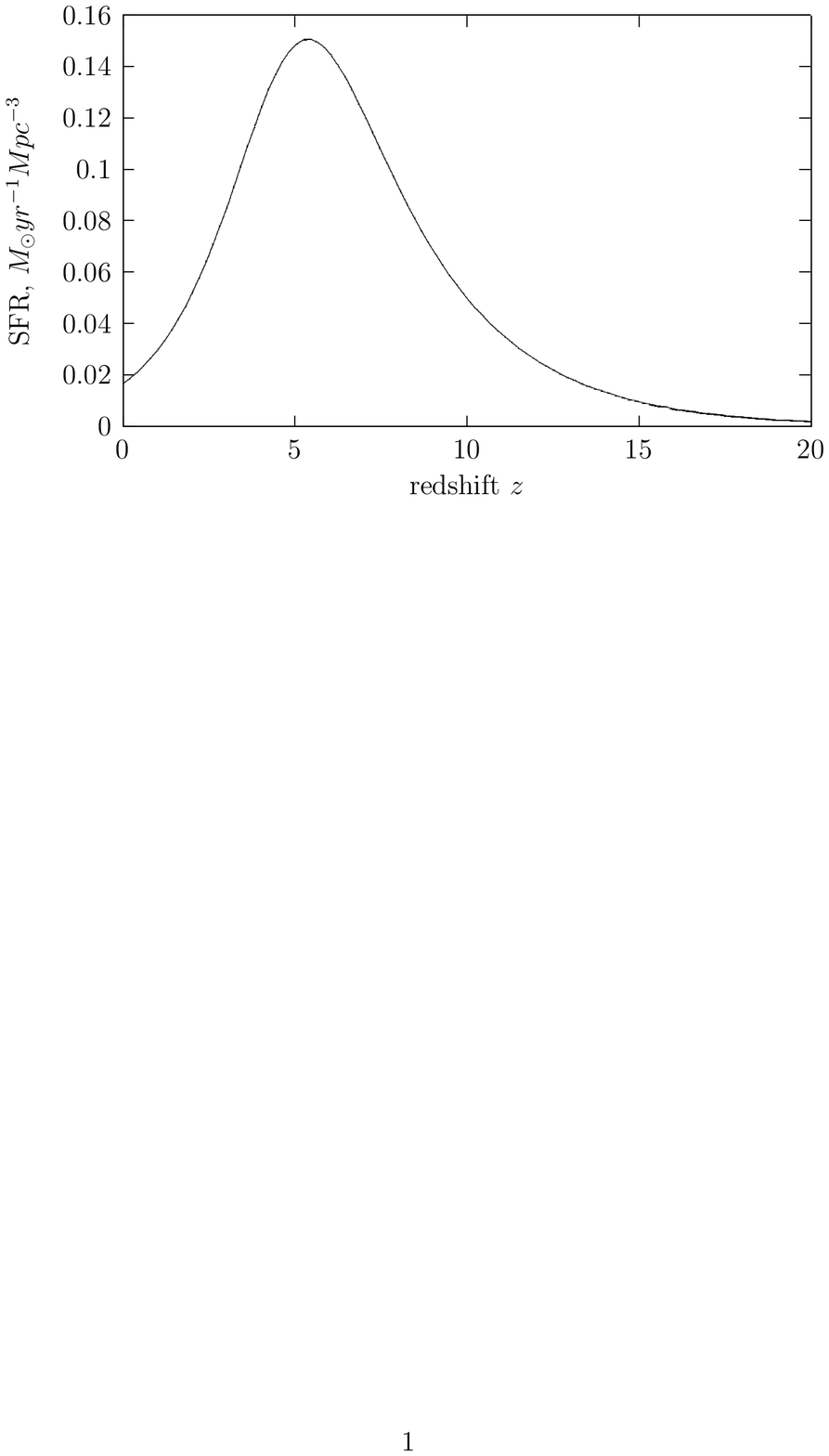}\\
(a)\\
\includegraphics{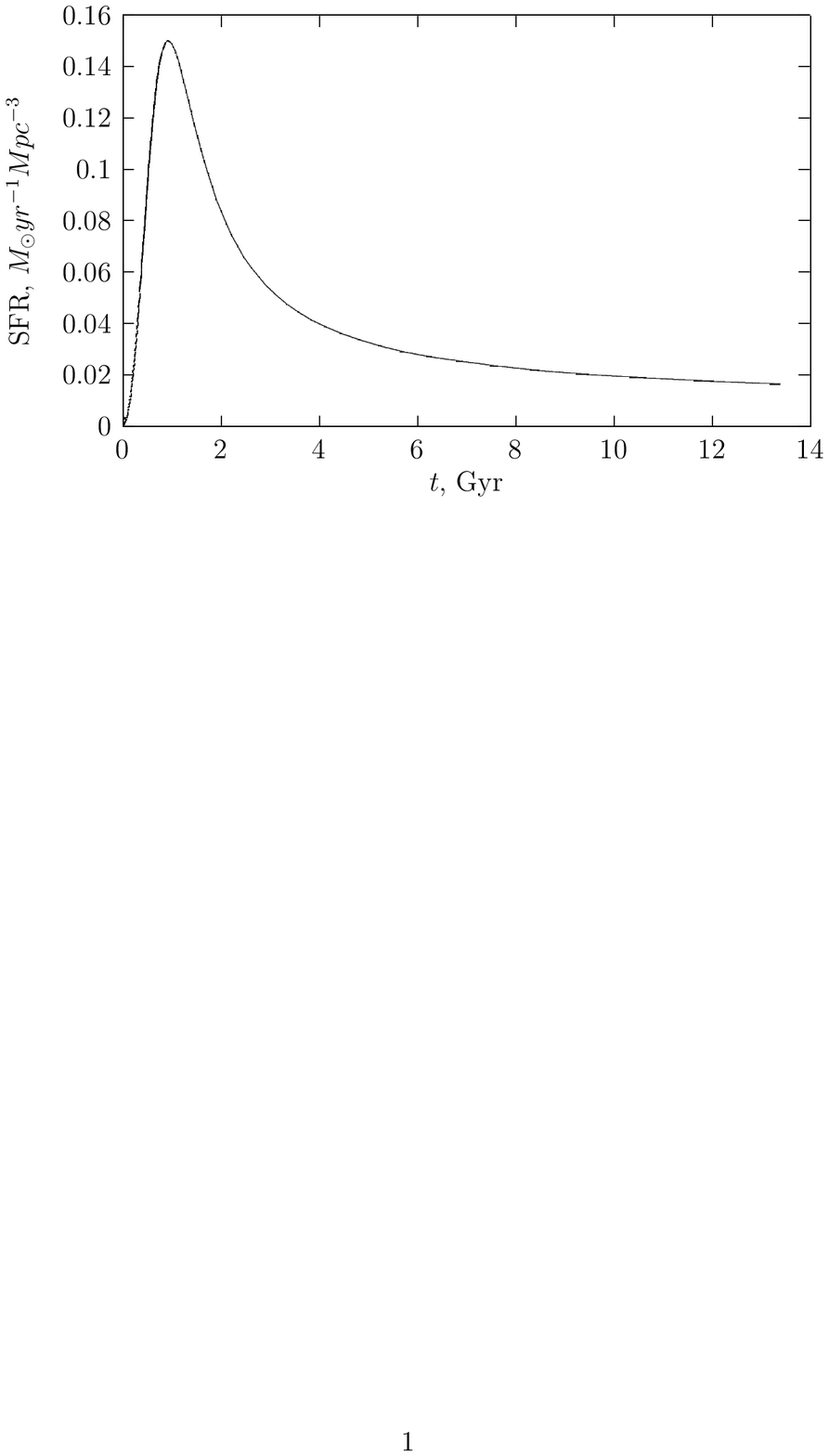}\\
(b)
  \\
  \caption{Star formation rate ($M_{\odot}yr^{-1}Mpc^{-3}$) as function of redshift (a) and time (b).}
  \end{center}
\end{figure}

Therefore we can estimate the anthropic factor for $0<\Lambda<10\Lambda_{0}$ using the time dependence of SFR in fig. 1(b). From (\ref{2}),(\ref{Vc}) it follows that
\begin{equation}
N(\Lambda)\sim \Lambda^{-1/2}\int_{0}^{t_{c}}d t \dot{n}(t)\left(\int_\tau^{\infty}\frac{d\tau}{\sinh^{2/3}(\tau)}\right)^{3}.
\label{antr}
\end{equation}
In Eq. (\ref{antr}) integration is over dimensionless time which is linked to our dimensionless time $\tau_{0}$ by the relation $\tau=\tau_{0}\sqrt{\Lambda/\Lambda_{0}}$. For simplicity, we assume that $\dot{n}(t)=0$ after  $t_{f}\sim 14$ Gyr. For our universe, $\tau_{f}=\tanh^{-1}\Omega_{\Lambda}$. This assumption is in a good agreement with VIMOS VLT Deep Survey data \cite{Star-7}. According to VVDS, SFR declines steadily by a factor 4 from $z=1.2$ to $z=0.05$, since in this phase both giant and intermediate galaxy populations decline. The most luminous sources ceased to efficiently produce new stars 12 Gyrs ago (at $z\sim4$), while intermediate luminosity sources
continued to produce stars till 2.5 Gyrs ago (at $z\sim0.2$).

It is convenient to define the relative anthropic factor as:
\begin{equation}
N_{rel}(\alpha)=\frac{N(\Lambda)}{N(\Lambda_{0})}\sim \alpha^{-1/2}\int_{0}^{t_{f}}d t \dot{n}(t)\left(\int_{t/T}^{\infty}d\tau\ f(\tau)\right)^{3}\times
\label{antr1}
\end{equation}
$$
\times\left[\int_{0}^{t_{f}}d t \dot{n}(t)\left(\int_{t/T_{0}}^{\infty}d\tau f(\tau)\right)^{3}\right]^{-1},
$$
where $\alpha=\Lambda/\Lambda_{0}$ and $f(\tau)=\sinh^{-2/3}(\tau)$.

Following \cite{Bousso}, one can assume that observers appeared in universe after a "delay time" of the order of some billions of years. One can re-write Eq. (\ref{2}) as
\begin{equation}
N(\Lambda)\sim \int^{t_{c}}_{\Delta t} \dot{n}(t-\Delta t, \Lambda)V_{c}(t)dt\sim \int^{t_{c}+\Delta t}_{0} \dot{n}(t, \Lambda)V_{c}(t+\Delta t)dt,
\label{22}
\end{equation}
where $\Delta t$ is a time delay. In this case in (\ref{antr1}) one need just to change the limit of the inner integral: $t\rightarrow (t+\Delta t)$.

The results of calculations for $\Delta t=0, \quad 5, \quad 10$ Gyr are given in figs. 2 - 4, respectively. {In the quoted figures, data of numerical calculations are marked by crosses, while solid lines are analytical fits.} {The result of the quoted figures is that the relative anthropic factor (and therefore the probability to find oneself in a universe with a given $\Lambda$) increases with decreasing $\Lambda$. The rapidity of this increasing is smallest for $\Delta t=0$ and largest for $\Delta t=10$ Gyr.}
\begin{figure}
\begin{center}
\includegraphics{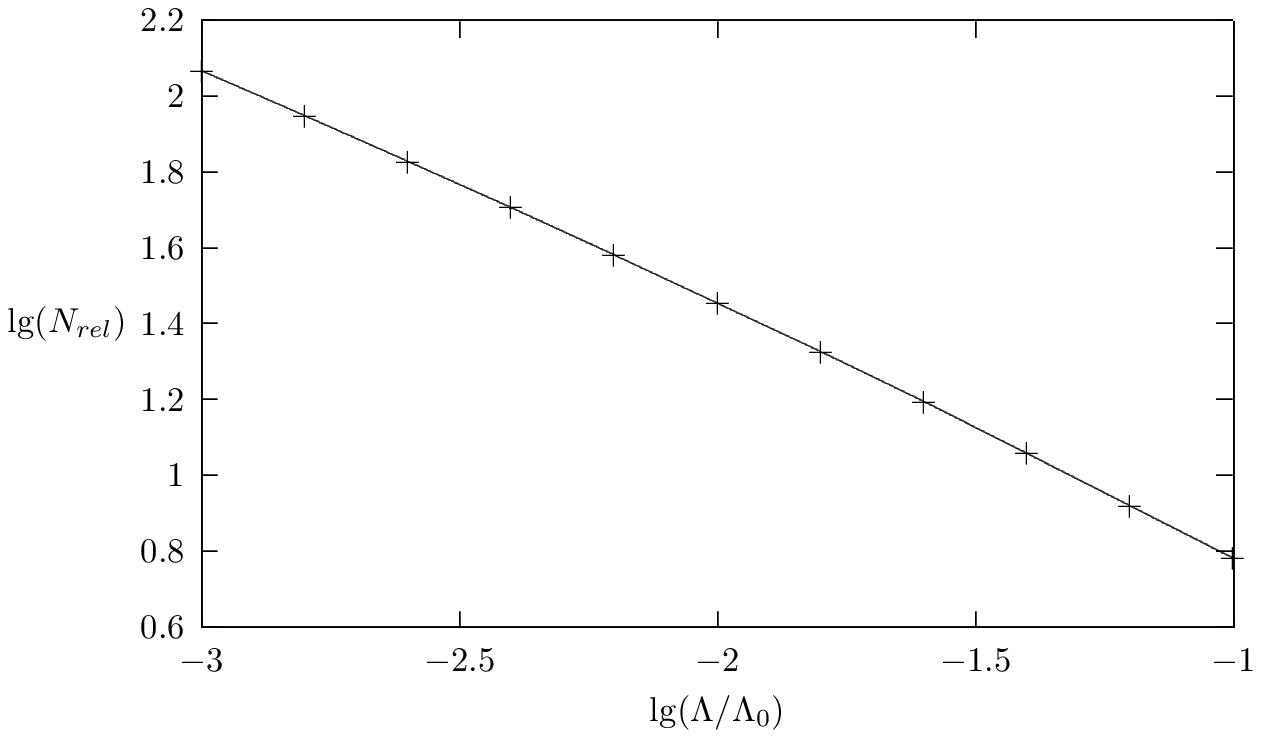}
\includegraphics{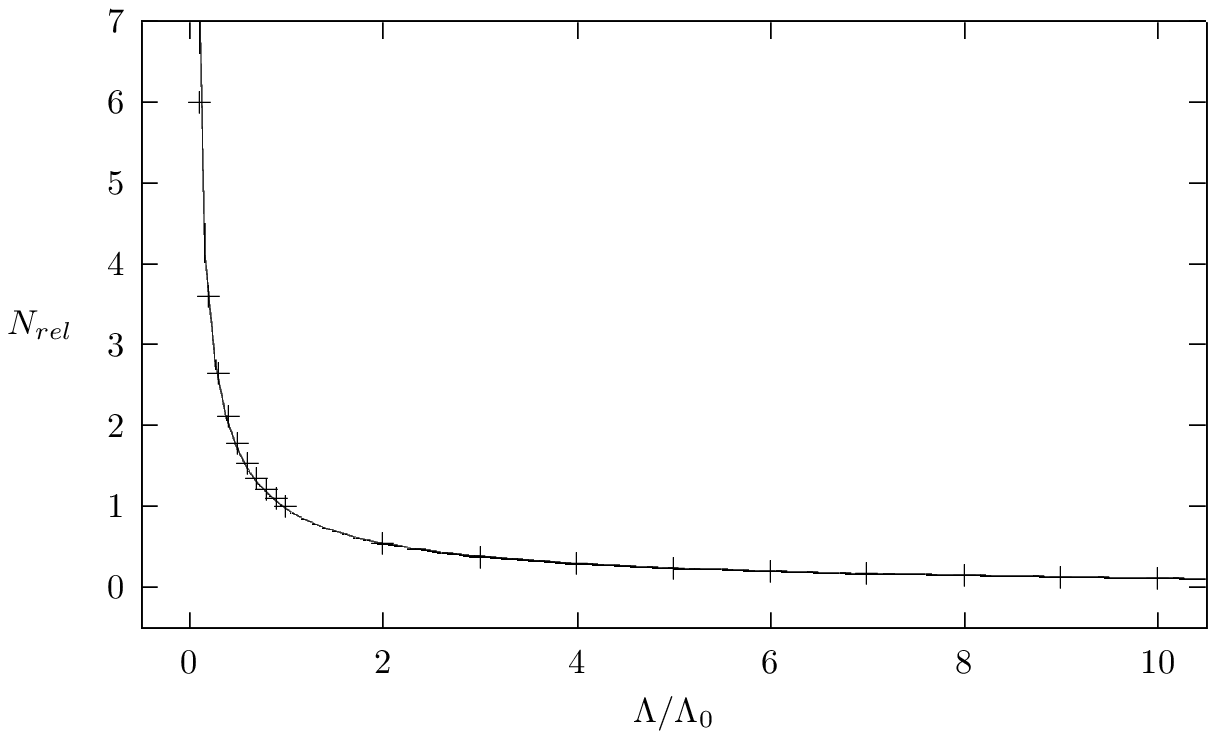}
  \\
  \caption{The relative anthropic factor for $\Delta t=0$. As shown in the upper panel, we use a logarithmic scale for small values of $\Lambda$.}
  \end{center}
\end{figure}
We found the following analytical fit for numerical estimations plotted in Figs. 2-4:
\begin{equation}
N_{rel}(\alpha) \approx \left\{\begin{array}{ll}  \label{Nrel}
\alpha^{-\gamma_{1}}\exp(-\alpha/\beta),\quad 0.1\Lambda_{0}\leq\Lambda\leq10\Lambda_{0}\\
\alpha^{-\gamma_{2}(\alpha)},\quad 0<\Lambda\leq 0.1\Lambda_{0}\end{array}\right.
\end{equation}
The parameters $\gamma_{1}$ and $\beta$ depend on $\Delta t$ (see table). Parameter $\gamma_{2}$ slowly decreases with decreasing $\alpha$. When $\Lambda\rightarrow 0$ the relative anthropic factor tends to
$$
N_{rel}(\alpha)\rightarrow\frac{C}{\alpha^{1/2}},\quad \alpha\rightarrow 0,
$$
where
$$
C=\int_{0}^{t
_{f}}d t \dot{n}(t)\left(\int_{0}^{\infty}d\tau\ f(\tau)\right)^{3}\left(\int_{0}^{t_{f}}d t \dot{n}(t)\left(\int_{t/T_{0}}^{\infty}d\tau f(\tau)\right)^{3}\right)^{-1}
$$
is a constant. At $\Lambda=0$, our calculations lose all meaning because in this case relative anthropic factor becomes infinitely large. Hence the probability that randomly selected observer measure a value $\Lambda=0$ is exactly equal to 1.
\begin{center}\begin{tabular}{|c|c|c|}
\hline
  $\Delta t$, Gyr & $\gamma_{1}$ & $\beta$  \\
\hline
  0 & 0.79 & 30 \\
  5 & 1.08 & 10 \\
  7.5 & 1.21  & 10 \\
  10 & 1.33 & 10\\
  12.5 & 1.45  & 10 \\
\hline
\end{tabular}
\end{center}

\begin{figure}
\begin{center}
\includegraphics{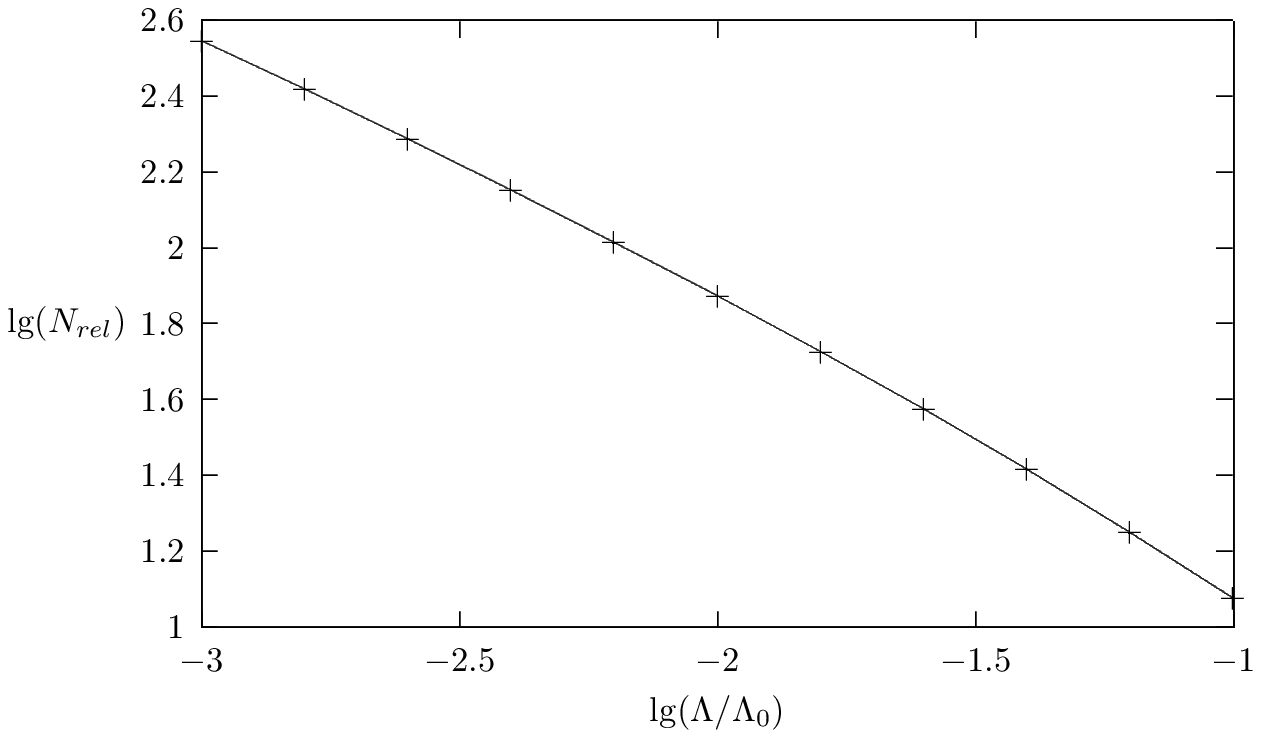}
\includegraphics{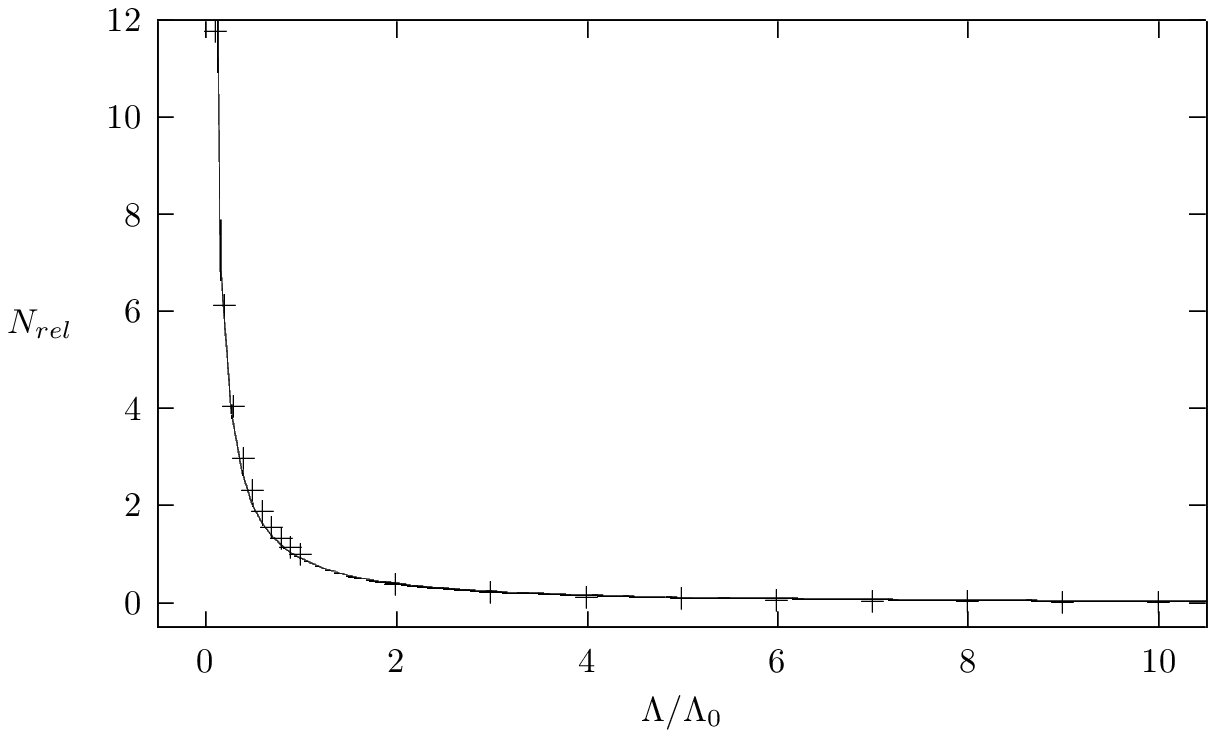}
  \caption{The relative anthropic factor for $\Delta t=5$ Gyr.}
\end{center}
\end{figure}

\begin{figure}
\begin{center}
\includegraphics{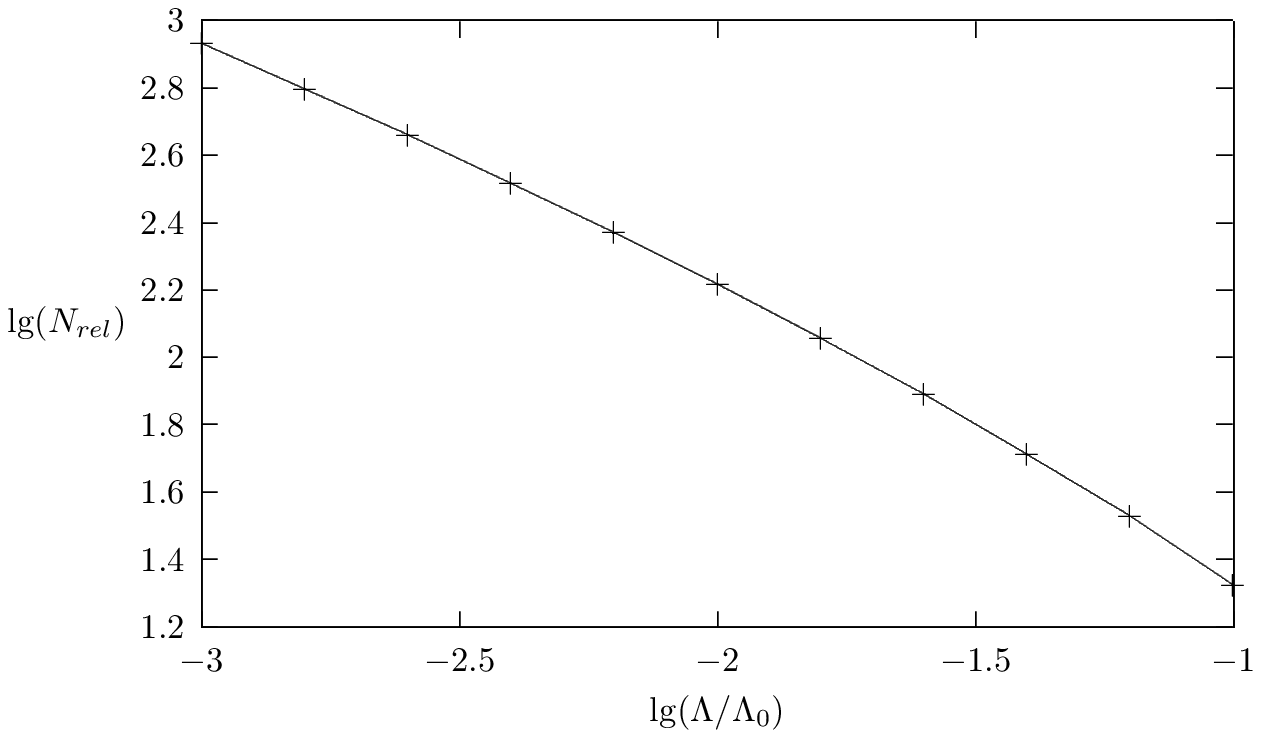}
\includegraphics{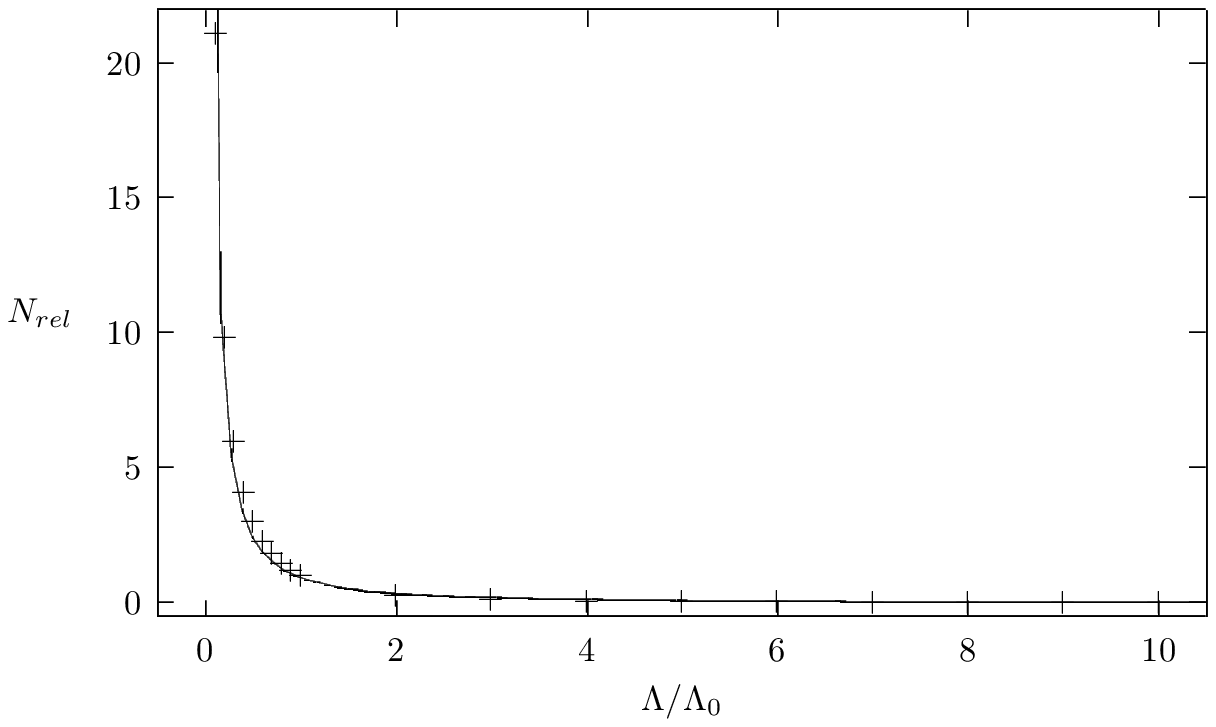}
\\
  \caption{The relative anthropic factor for $\Delta t=10$ Gyr.}\
\end{center}
\end{figure}

\section{The 4-volume averaging of probabilities as a possible solution to the infrared divergence problem}

If the eternal inflation model is correct, this implies the existence of multiverse, containing infinite number of copies of every possible observer. What can we tell about probabilities of a event in such a multiverse? Suppose we are conducting an experiment to determine the value of cosmological constant with various possible outcomes $\Lambda_{n}$. The probabilities for {these outcomes} are connected by the formula
\begin{equation}
\frac{P_{\Lambda_{n}}}{P_{\Lambda_{k}}}=\frac{N(\Lambda_{n})}{N(\Lambda_{k})},
\label{Prob}
\end{equation}
where $N(\Lambda_{n, k})$ is the amount of $\Lambda_{n}$ and $\Lambda_{k}$ results in all the multiverse.
The direct calculation of probabilities by Eq. (\ref{Prob}) becomes impossible because $N(\Lambda_{k,n})\rightarrow\infty$.
The task of infinities elimination, known as "measure problem", is important for modern cosmology. Some other possibilities for the measure definition have been proposed so far \cite{Measure-1}-\cite{Measure-9}.

Among the various approaches to measure problem, one should notice the one described in \cite{Page-1}. The key idea, in \cite{Page-1}, is that one need to replace volume weighting of probabilities by volume averaging. {According to this assumption the relative probabilities are proportional to expectation values of the fraction of the number of locations in which the observation
occurs.} The latter value is proportional to a number of observers per unit 4-volume, i.e. the number of occurred observations per unit spatial volume per unit of time. Therefore one need to compare the densities of observers instead their numbers.  

Initially, volume averaging was introduced in order to solve the Boltzmann brain (BB) problem \cite{Albrecht-1}, \cite{Albrecht-2}. {This problem can be described as follows. Let`s consider toy multiverse containing only two types of universe. The universe of first type (I) expands forever while the universe of second type (II) has a finite size and finite lifetime. In I-universes , ``ordinary observers'', like ourselves, should be vastly outnumbered by infinite number of BBs, arising from vacuum fluctuations. In II-universes only a finite number of ordinary observers exists, therefore the ''ordinary observers'' in such multiverse are highly atypical. As pointed in \cite{Page-1}  replacing the volume weighting measure with volume averaging can avoid the BB catastrophe because the density of BBs is much less than the density of ordinary observers.} Volume averaging also has a deep link with quantum mechanics \cite{Page-2}, \cite{Page-3}, \cite{Page-4}.

One can show that volume averaging eliminates the "infrared divergence". The spatial volume corresponding to comoving volume in (\ref{Vc}) is
\begin{equation}
V_{3}=\frac{4\pi T^{3}}{3}\left(\sinh^{2/3}(\tau)\int^{\infty}_{\tau}\frac{d\tau}{\sinh^{2/3}(\tau)}\right)^{3}
\end{equation}
Obviously, for $\tau\rightarrow\infty$, $V_{3}$ converges to $4\pi(\sqrt{3/\Lambda})^3/3$, i.e. to the Hubble volume of bubble.
The 4-volume therefore is
$$
V_{4}(t)=\int_{0}^{t}V_{3}dt=\frac{4\pi T^{4}}{3}\int_{0}^{\tau}d\tau \left(\sinh^{2/3}(\tau)\int^{\infty}_{\tau}\frac{d\tau}{\sinh^{2/3}(\tau)}\right)^{3}
$$
and diverges at $\tau\rightarrow\infty$. The density of observers per unit 4-volume tends to zero for long-lived de Sitter vacua, but the relative density of observers is non-zero
\begin{equation}
\langle N_{rel}\rangle=\frac{N(\Lambda)/V_{4}(\Lambda)}{N(\Lambda_{0})/V_{4}(\Lambda_{0})}=\alpha^{2}N_{rel}.
\label{aver}
\end{equation}
So the probability to find oneself in a universe with given value of $\Lambda$ becomes a well-defined function of $\Lambda$. Combining (\ref{Nrel}) and (\ref{aver}) gives the following result
\begin{equation}
\langle N_{rel}(\alpha)\rangle\approx \left\{\begin{array}{ll} \label{Naver}
 \alpha^{2-\gamma_{1}}\exp(-\alpha/\beta),\quad 0.1\Lambda_{0}\leq\Lambda\leq10\Lambda_{0}\\
 \alpha^{2-\gamma_{2}(\alpha)},\quad 0<\Lambda\leq 0.1\Lambda_{0}\\
  \end{array}\right.
\end{equation}
{The dependence of $N_{rel}$ on $\Lambda$ is depicted on Fig. 5 for $\Delta t=5$ (thin solid line), 7.5 (thin dotted line), 10 (thick solid line), and 12.5 Gyr (thick dotted line).} Hence the density of observers reaches its maximum for a value $\Lambda_{m}$ of the vacuum energy. {The value of this maximum decreases with increasing time delay.}
For a time delay 2.5 Gyr$<\Delta t<$12.5 Gyr, we have $\Lambda_{m}\approx5\div10 \Lambda_{0}$. For larger values of vacuum energy, $\Lambda>10\Lambda_{0}$ one has to take into account the star formation decline due to early dominance of vacuum energy.

Finally, it remains to show that volume averaging does not lead to singularities in the case $\Lambda=0$. Let`s consider the unit comoving volume in such universe. The total stellar mass within this unit volume can be estimated as
$$
M^{(1)}=\int_{0}^{t_{f}}\dot{n}(t)dt.
$$
The corresponding 4-volume increases with time according to the law
$$
V^{(1)}_{4}(t)=\int^{t}_{0} t^{2}dt=t^{3}/2.
$$
Hence the density of observers is equal to
$$
\langle N(\Lambda=0)\rangle\sim\lim_{t\rightarrow\infty}\frac{M^{(1)}}{t^{3}/2}
$$
and it is easy to see that
$$
\langle N_{rel}(0)\rangle\sim\lim_{t\rightarrow\infty}\frac{\int_{0}^{t/T_{0}}d\tau \left(\sinh^{2/3}(\tau)\int^{\infty}_{\tau}\frac{d\tau}{\sinh^{2/3}(\tau)}\right.)^{3}}{t^{3}}=0
$$
So the point $\Lambda=0$ is not singular.
\begin{figure}
\begin{center}
\includegraphics{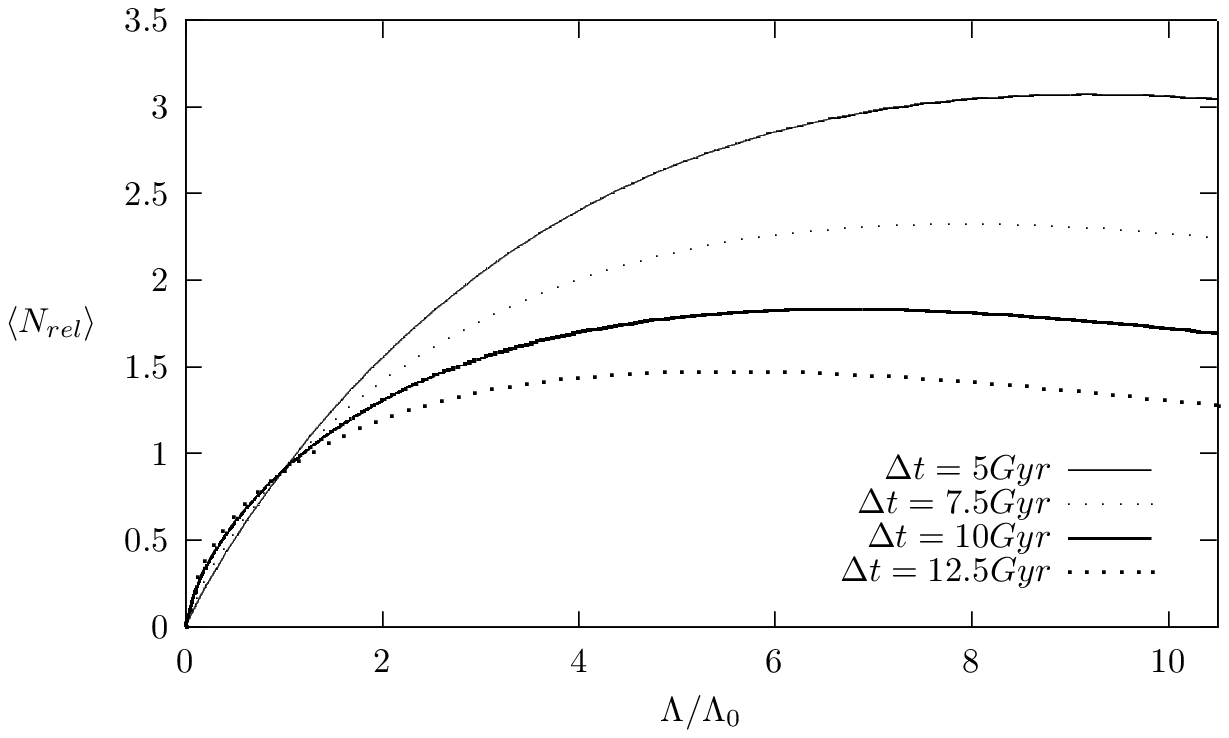}
\\
\caption{The relative density of observers for various $\Delta t$}
\end{center}
\end{figure}

\section{Conclusion}

Some questions remain to be answered. They are the following:

1. the observable vacuum energy density is one order of magnitude smaller than the "optimal" value $\Lambda_{m}$. Does this mean that our universe is atypical in the multiverse?

The traditional approach is that the observable universe is assumed a typical one. It seems to us that this methodology only complicates the understanding of the real universe. In our opinion, the knowledge of fundamental laws is enough to estimate parameters of a "typical" (from anthropic point of view) universe. Subsequent comparison of these parameters with the observed values in our universe help us to find an answer to the aforementioned question. One note also that according to our calculations (fig. 5) the probability to find oneself in the universe with $\Lambda=\Lambda_{m}$ is only 1.5 - 3 times higher than probability to find oneself in the universe with $\Lambda=\Lambda_{0}$. Therefore the observed value of vacuum energy lies in the reasonable region.

2. Negative values of $\Lambda$. One can notice that, for $\Lambda<0$, the volume averaging also eliminates "infrared divergence" at $|\Lambda|\rightarrow0$. Hence regularization scheme based on volume averaging gives correct answers for $\Lambda<0$.
But the following problem appears. If vacuum energy is negative, universe ends its existence in big crunch singularity. In this case the density of observers is larger than in the case of a de Sitter universe, because the 4-volume of anti de Sitter universe is much less. So, why we don't live in universe with negative $\Lambda$?

Firstly, this problem occurs in the \textit{prediction} stage, but it disappears at the \textit{explanatory} stage when cosmological constant has already been measured by the observer. According to \cite{Vilenkin}, only the observers with similar informational content can be assigned to the same equivalence class. The sign of the cosmological constant is already known for us. Hence we are representatives of the reference class which includes all observers for which $\Lambda>0$. When calculating probabilities one should consider only observers belonging to this class.

Secondly, there may exists a physical mechanism, unknown to date, imposing the bounds on the lifetime (and 4-volume) of de Sitter universe \cite{Doomsday-1}, \cite{Doomsday-2}. An interesting scenario was suggested in \cite{Page-5}, \cite{Page-6}, \cite{Page-7}. According to this scenario our vacuum should be rather unstable and should decay within 20 Gyr (which is possible if the gravitino is superheavy).

In conclusion it should be emphasized that the main result of the present paper is the proof that replacing volume weighting with volume averaging in the cosmological measure, can avoid the infrared divergence problem. Volume averaging leads to natural explanation of the reason why the observed value of vacuum energy is non-zero. Perhaps this result can be considered as argument in favor of using volume averaging measure in cosmology.

\acknowledgments

A. A. grateful to A. V. Yurov for useful discussion. Authors also grateful to anonymous referees for useful comments.

\end{document}